\documentclass[conference]{IEEEtran}

\IEEEoverridecommandlockouts

\usepackage{amsmath,amssymb,amsfonts}   
\usepackage{graphicx}                   
\usepackage{textcomp}                   
\usepackage{xcolor}                     
\usepackage{longtable}                  
\usepackage{multirow}                   
\usepackage{soul}                       

\usepackage{algorithm2e}                
\usepackage{algpseudocode}              

\usepackage{hyperref}                   

\usepackage{bm}                         

\def\BibTeX{{\rm B\kern-.05em{\sc i\kern-.025em b}\kern-.08em T\kern-.1667em\lower.7ex\hbox{E}\kern-.125emX}}

\makeatletter
\def\ps@IEEEtitlepagestyle{%
  \def\@oddfoot{\mycopyrightnotice}%
  \def\@oddhead{\hbox{}\@IEEEheaderstyle\leftmark\hfil\thepage}\relax
  \def\@evenhead{\@IEEEheaderstyle\thepage\hfil\leftmark\hbox{}}\relax
  \def\@evenfoot{}%
}
\def\mycopyrightnotice{%
  \begin{minipage}{\textwidth}
  \centering \scriptsize
  Copyright~\copyright~2024 IEEE. Personal use of this material is permitted. Permission from IEEE must be obtained for all other uses, in any current or future media, including\\reprinting/republishing this material for advertising or promotional purposes, creating new collective works, for resale or redistribution to servers or lists, or reuse of any copyrighted component of this work in other works.
  \end{minipage}
}
\makeatother

\begin{document}

\title{Uncertainty Quantified Deep Learning and Regression Analysis Framework for Image Segmentation of Skin Cancer Lesions\\
{\footnotesize \textsuperscript{}}
\thanks{}
}

\author{
    \IEEEauthorblockN{1\textsuperscript{st} Elhoucine Elfatimi}
    \IEEEauthorblockA{\textit{Department of Pathology Laboratory Medicine} \\
    \textit{University of California}\\
    Irvine, CA, USA \\
    eelfatim@uci.edu}
    \and
    \IEEEauthorblockN{2\textsuperscript{nd} Pratik Shah*\thanks{*Corresponding author: Dr. Pratik Shah Ph.D. (pratik.shah@uci.edu)}}
    \IEEEauthorblockA{\textit{Department of Pathology Laboratory Medicine} \\
    \textit{Biomedical Engineering} \\
    \textit{University of California}\\
    Irvine, CA, USA \\
    pratik.shah@uci.edu}
}

\maketitle

\begin{abstract}
Deep learning models (DLMs) frequently achieve
accurate segmentation and classification of tumors from medical images. However, DLMs lacking feedback on their image seg-
mentation mechanisms such as Dice coefficients and confidence in their performance face challenges processing previously unseen
images in real-world clinical settings. Uncertainty estimates to identify DLM predictions at the cellular or single-pixel levels
requiring clinician review can enhance trust, however their deployment requires significant computational resources. This
study reports two DLMs, one trained from scratch and another based on transfer learning, with Monte Carlo dropout or Bayes-
by-backprop uncertainty estimations to segment lesions from the publicly available The International Skin Imaging Collaboration-
19 dermoscopy image database with lesions of cancer. A novel approach to compute pixel-by-pixel uncertainty estimations of
DLM segmentation performance in multiple clinical regions from a single dermatoscopy image with corresponding Dice scores is reported for the first time. Image-level uncertainty maps demonstrated correspondence between imperfect DLM segmen-
tation and high uncertainty levels in specific skin tissue regions with or without lesions. Four new linear regression models that can predict the Dice performance of DLM segmentation using
constants and uncertainty measures either individually or in
combination from lesions, tissue structures, and non-tissue pixels
regions critical for clinical diagnosis and prognostication in skin
images (Spearman’s correlation, p $< $0.05), are reported for the
first time for low-compute uncertainty estimation workflows.

\end{abstract}

\section{Introduction}
\label{sec:intro}

Skin cancer is the most common cancer in the United States, affecting about one in five Americans in their lifetime ~\cite{SkinCancerFoundation}. Differentiating between melanoma, melanocytic nevus (mole), and seborrheic keratosis can be clinically challenging. Seborrheic keratosis appears as discolored, raised patches and is unrelated to sun exposure, while melanoma, an aggressive skin cancer, is caused by UV light damaging skin cell DNA ~\cite{aad}. Melanocytic nevi are benign skin lesions from melanocytes, generally harmless, making visual inspection for skin cancer difficult.
Deep neural networks (DNNs) are employed for image detection, classification, and segmentation in clinical applications. Research has focused on using deep learning algorithms to classify and segment dermoscopic images of skin lesions ~\cite{goyal2020artificial}. However,  DNNs often lack mechanisms for risk evaluation and comprehensive uncertainty estimations, which are essential for assessing the reliability and confidence of model predictions ~\cite{risk-analysis}.  Many DNNs use softmax activation, which can lead to unjustified confidence levels for unknown inputs ~\cite{pmlr-v48-gal16}. Bayesian neural networks that provide output and uncertainty by setting a distribution over model weights, are computationally complex and impractical for high-dimensional inputs ~\cite{BISHOP1997}. {Approximating Bayesian distributions using dropout-trained networks can address these challenges} ~\cite{pmlr-v48-gal16}.

One key goal of this study was to address clinically relevant challenges of uncertainty estimation in DNNs by employing Bayesian uncertainty methods-inspired approaches of Monte Carlo dropout (MCD) and Bayes-by-backprop (BBP) for segmentation of melanoma, melanocytic nevus, and seborrheic keratosis skin conditions from the The International Skin Imaging Collaboration (1SIC) 2019 challenge images\cite{tschandl2018ham10000}.   The most prevalent DNNs models do not to quantify prediction uncertainty at the pixel level, which limits their clinical applicability where prediction confidence is crucial. This study integrates MCD and BBP to provide pixel-level uncertainty estimates, thereby enhancing both the trustworthiness and interpretability of predictions. Another goal was approximating Bayesian probability distributions in a more computationally feasible manner, making them suitable for high-dimensional dermoscopy images with multiple classes. Additionally, deep learning models (DLMs) trained from scratch (TFS) and based on transfer learning (TL) were evaluated to compare their performance with MCD and BBP techniques. A novel fast computing framework that integrates multiple clinical skin region-based pixel-by-pixel uncertainty from a single image to predict DLMs output uncertainties using linear regression models is reported for the first time to visualize, interpret and improve DNN predictions for clinical applications.

\subsection{Related Work}
The relationship between Dice scores and uncertainty in deep learning models (DLMs) for segmenting clinically important regions in skin and medical images is not fully understood, and face challenges such as lack of voxel-level uncertainty feedback and methods for individualized clinical outcomes. A single Dice score is commonly used in organ segmentation studies. For instance, Senapati et al. evaluated liver segmentation using Bayesian neural networks with Dice scores \cite{senapati2020bayesian}, Hoebel et al. used linear regression with uncertainty metrics for thorax segmentation \cite{hoebel2020exploration}, and Do et al. demonstrated a correlation between Dice scores and uncertainty in myocardial imaging with a UNet-like architecture \cite{do2020accuracy}. Most studies on dermoscopic images use DLMs for classification without regression analysis or Dice score evaluations. For example, uncertainty distributions from MCD have been reported without Dice scores \cite{skin-ham10000}.Similarly, Combalia et al. used Monte Carlo sampling without computing Dice scores \cite{combalia2020uncertainty}, and Shapna employed a CNN model without pixel-level analysis \cite{SHAPNA}. In skin lesion segmentation, studies like Ashraf et al. used test-time augmentation and Bayesian learning to evaluate Dice coefficients but did not include linear regression \cite{ashraf2022melanoma}. Similarly, De Angelo et al. used Dice and Jaccard index for multiple regions of interests (ROIs) without linear regression \cite{de2019skin}. Another study reported improved U-Net architectures for skin segmentation on the ISIC dataset, but did not focus on uncertainty estimation using training from scratch (TFS) or transfer learning (TL) with MCD and BBP approaches used in our paper \cite{nawaz2022melanoma}. There is a need for a framework that evaluates the relationship between Dice scores, uncertainty, and the performance of DLMs trained with TFS or TL using MCD and BBP, particularly for clinical diagnosis of various skin lesions.


\subsection{Summary of Contributions}

This study focuses on training multiple DNNs using MCD and BBP techniques with TFS and TL approaches for segmentation of melanoma, melanocytic nevus, and seborrheic keratosis skin conditions from the ISIC-19 images and estimate DLM uncertainty at the pixel level. Another key goal was to enhance the segmentation accuracy and reliability of binary segmentation tasks, particularly in identifying skin or cancer lesions, from multiple clinically important ROI’s from dermoscopic Red Green Blue (RGB) images. The performance of these approaches was assessed using five-fold cross-validation and other comprehensive statistical metrics reported in the methods section and summarized in Tables \ref{tab1} and \ref{tab2}. By leveraging ground-truth skin lesion labels, all images were categorized into two clinical ROIs: pixels with lesions and pixels without skin lesions. Additionally, three individual clinical region-based uncertainties and an overall uncertainty measure for each image were computed specific to each of the three skin lesion diagnosis classes. A novel algorithm (Algorithm \ref{seg_algo1})is proposed to calculate region-specific uncertainty, facilitating a more granular evaluation of DLM performance. Four separate linear models were trained and validated to learn the correspondence between the Dice coefficient, calculated by comparing DLM output with ground truth segmentation masks for each image, and the clinical region-specific uncertainties derived from the MCD and BBP approaches (Algorithm \ref{seg_algo1}). Statistical analysis revealed significant correlations (p $<$ 0.05) between Dice scores and the predictive performance of the MCD and BBP DLMs. The linear models reported in this study demonstrated high-accuracy in predicting the Dice performance of DLMs using constants and uncertainty measures from two clinical ROIs, either individually or in combination (Equations \ref{eq:1} and \ref{eq:4}). 



\section{Data Description}

The International Skin Imaging foundation released a data set in 2019 (ISIC 2019) containing 13,786 3-channel RGB images of resolution ranging from (7000 $\times$ 4500) to (1024 $\times$ 720) (W $\times$ H) with available ground truth (GT) binary mask associated with each image as its label~\cite{ref-url1}. Each binary mask had white pixels denoting regions of the image where skin moles and or cancer labels were present (class 1). Black pixels in the binary masks represented non-target and or non-disease regions (class 0). Each image also has a clinical class (from a set of 17 classes) associated with it based on the dermoscopy diagnosis. The dataset used comprised 2,935 images with the top three available clinical classes: nevus (n=1497), melanoma (n=1019),and seborrheic keratosis (n=419). These classes are highly prevalent in clinical practice, making them a good fit for research and diagnostic algorithms. This dataset has been widely used  to identify the most common types of skin cancer from dermoscopic images \cite{combalia2019bcn20000}. Detailed description on data pre-processing and training of DLMS with five-fold cross validation were performed as previously described \cite {ghosal2021deep}.

\section{Methods}



\subsection{Deep Learning Model Training}

\newcommand{\inlineTilde}{\raisebox{-0.5ex}{\texttt{\char`~}}}

The ISIC 2019 dataset was split into 80\% for training (\inlineTilde 2,347 images) and 20\% for testing (\inlineTilde588 images).A Bayesian VGG-UNet CNN, enhanced with MCD and BBP techniques, was trained on skin lesion images and their labels. The network used an encoder-decoder structure \cite{ghosal2021deep}, based on a modified VGG-16 \cite{Simonyan2015VeryDC}. To address uncertainties in segmentation, uncertainty quantification mechanisms were integrated into the decoder component to mitigate noise during the downscaling phase. The MCD and BBP approximated Bayesian inference, minimizing the approximate posterior within a deep Gaussian process framework \cite{pmlr-v48-gal16}.  All DLMs were trained for 50 epochs with a batch size of two using the Adam optimizer and a learning rate of ($\alpha_{lr} = 0.001$). Input images were resized to 608 × 416 (width × height), with outputs at 304 × 208. Training was done on an NVIDIA GeForce RTX 4080 Ti with 32GB of GPU memory.

\begin{figure}[h!]
\centering
\includegraphics[width=0.48\textwidth, height=0.23\textheight]{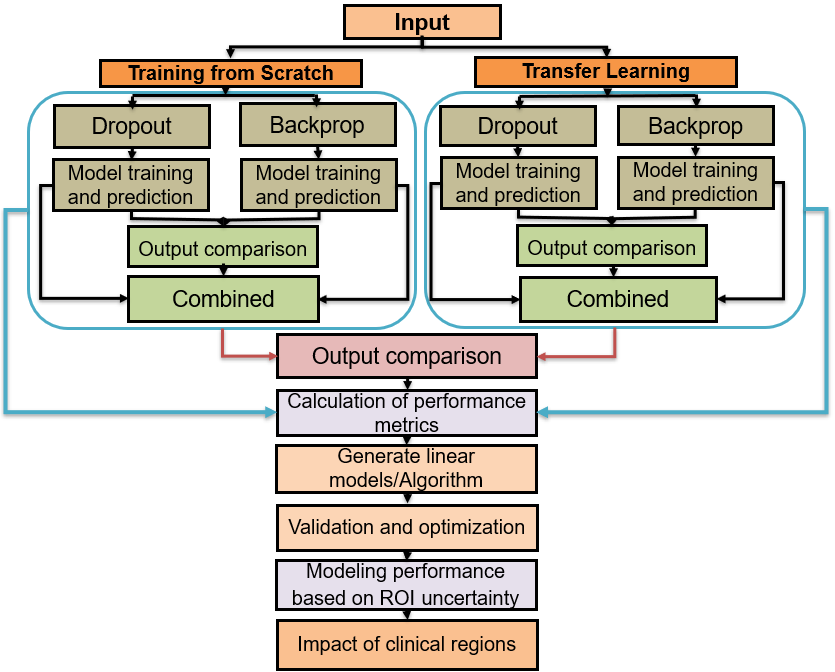}
\caption{Schematic diagram of the deep learning framework for interpretable skin lesion image segmentation with uncertainty quantification and linear regression }
\label{fig1}
\end{figure}

The DLM training, implementation, and image processing workflow, shown in Fig.\ref{fig1}, involved training using both TL and TFS. Outputs were generated from DLMs with Monte Carlo Dropout (DLM MCD) and Bayes-by-Backprop (DLM Bc). The output predictions from DLM MCD and DLM Bc were combined and compared with ground truth segmentation masks, and Dice scores were computed. Correlation analysis was used to establish the relationship between Dice scores and uncertainty estimations derived from the DLM by developing separate linear models and algorithms for predictions that incorporated uncertainties. This framework integrates the segmentation performance of multiple DLMs with specific regions of clinical relevance in individual skin images to calculate the overall segmentation performance of the deep learning model.

\vspace{-0.3cm}
\vspace{-0.4cm}
\subsection{Deep Learning Model Prediction And Evaluation}\label{sec_model_pred_eval}

For each model prediction, the numbers of True Positive (TP), False Positive (FP), False Negative (FN), and True Negative (TN) pixels were counted over the entire output. A TP instance was when an image lesion pixel was segmented accurately, and TN when a non-lesion or non-tissue pixel indicated in ground truth clinical labels was not segmented by the deep learning models. False negative pixels were those that were denoted as lesion pixels in the ground truth, but were not segmented by the model, and FP pixels were those segmented as a lesion by the model but classified as non-lesion tissue or non-tissue in the ground truth data. Using these definitions, four metrics for model predictions for all the images in the test data - the True Positive Rate (TPR), False Positive Rate (FPR) ( Table~\ref{tab1}), and the AUROC and Dice scores that evaluated the similarity between the ground truth clinical labels and the DLM generated segmentation were calculated (Table\ref{tab1}). The Dice scores were used for subsequent linear model estimations. The overall median Dice, AUROC, TPR and FPR values for all 2,935 test images are reported in Table~\ref{tab1}.  A comprehensive comparative analysis of the uncertainty from TFS and TL backprop vs. dropout DLM was learned with linear regression for predicting the Dice coefficient of skin lesion and melanoma segmentation. For each image, a distribution of pixel-level uncertainty was obtained. The  overall mean uncertainty and other  uncertainty metrics for all 2,935 test images are reported in Table~\ref{tab2}. Uncertainty estimates resulting from this evaluation range between a minimum value of 0 for the model being fully certain of its prediction to a maximum of 1 when the model is fully uncertain about its prediction. The analysis was expanded to explore the combined effect of MCD and Bc approaches within the TFS and TL frameworks (Fig.~\ref{fig2}). By integrating these techniques, we aimed to harness the synergistic benefits they offer in mitigating uncertainty. To obtain pixel-level uncertainty, 50 MC iterations were taken where the difference between 67\textsuperscript{th} and 33\textsuperscript{rd} percentiles of their sigmoid values from the DLMs predictions were set as uncertainty estimates and represented as heatmaps (see Fig.~\ref{fig4} and Fig.~\ref{fig3}). Two hundred and ninety-four images containing lesion labels from the test dataset constituted the nevus class subset, 224 images from the test data containing lesion label constituted the melanoma class subset, and 87 images that contained seborrhoeic keratosis lesion labels from the test data set were used.
Mathematically, for a given DL MCD model $M$ and input test image $X_{test_i}$, the DLM prediction was calculated by: $O_{M_{p}}^{j}$ = argmax($M_{p}$($X_{test_i}$)) for the $j$\textsuperscript{th} MC iteration, where $M_{p}$ is the model $M$ in testing phase. The final model output $\bm{O_{M_i}}$ was generated by taking a mean of all 50 MC iteration outputs. To generate the predicted segmentation outputs, a value based thresholding was used to designate the predicted class to which the pixel belonged. Predicted ‘class 0' (pixel value of ‘0' in binary scale) or non-lesion pixels were assigned a value of 0 only if the mean value for the corresponding pixel across all MC iterations was $\leq 0.95$, and predicted ‘class 1' (pixel value of ‘1' in binary scale) or lesion pixels as 1 were assigned only if the mean value from the MC iterations was $> 0.95$. Thus, 
\begin{equation}
    O_{M_{p1}} = 1/\alpha \sum_{j=1}^{\alpha} O_{M_{p}}^{j}
\end{equation}
where $\alpha$ = 50 is the number of MC iterations. And the final model output, $\bm{O_{M_i}} = [O_{M_{p1}}$ s.t $\forall$ pixels, $p$ in $O_{M_{p1}}$, ($\lvert p \rvert$ $>$ 0.95) $\implies$ ($\lvert p \rvert$ = 1) and ($\lvert p \rvert$ $\leq$ 0.95) $\implies$ ($\lvert p \rvert$ = 0)]. To obtain the corresponding uncertainty map, $\bm{Unc_{map_i}}$,  each MC iteration step produced an uncertainty map as $Unc_{M_{p}}^{j}$ = max($M_{p}$($X_{test_i}$)). Thus,
\begin{equation}
    \bm{Unc_{map_i}} = \bm{P_{67}}([Unc_{M_{p}}^{j}]_{j=1}^{\alpha}) - \bm{P_{33}}([Unc_{M_{p}}^{j}]_{j=1}^{\alpha})
\end{equation}



\subsection{Modelling Performance Based on Uncertainty} \label{sec:results_and_disc3}

Algorithm~\ref{seg_algo1} is proposed to calculate the estimated Dice performance of DLMs using uncertainties for clinical ROIs in skin images used in this study. Equation~\ref{eq:1} calculated the Dice of the segmentation model based on a linear combination of lesion-tissue and non-lesion skin pixels uncertainties, and equations \ref{eq:4}(i) and \ref{eq:4}(ii) calculated the Dice of the DLM segmentation from the pixels of three individual ROIs. Two subsets of test data were used for generating the linear regression models and for obtaining the values of the coefficients for equations~\ref{eq:1} to~\ref{eq:3}. A detailed breakdown of the coefficients and associated values that explain the uncertainty within specific ROIs for three classes are in Table \ref{combined_results_horizontal}. The predictive capabilities of the proposed Algorithm 1 and linear models are further described in subsection ~\ref{subsection:predictive performance}.
\begin{algorithm}[t!]
\small
\caption{\textbf{Uncertainty Estimations of Clinical Regions of Interest (ROI)}: Predicting deep learning model (DLM) performance  with uncertainty estimations of clinical class-based image regions}\label{seg_algo1}
\hrule
\vspace{1mm}

\textit{\textbf{Step 1: DLM Predictions and Uncertainty Mapping}}\\
\hspace*{5mm} \textit{\textbf{Input:}} Optimized deep learning model $M$, test dataset $X_{test}$ with $n$ images, and corresponding labels $X_{GT}$.
\newline
\hspace*{5mm} \textit{\textbf{Output:}} DLM segmentation output, $\bm{O_{M_i}}$ and uncertainty map, $\bm{Unc_{map_i}}$ (see methods Section~\ref{sec_model_pred_eval}) corresponding to $X_{test_i}$.
Thus, $\forall$ i = [1, 2,....., $n$] images, 
\[
X_{Dice_i} = Dice(O_{M_i}, X_{GT_i}), List_{Dice} = [X_{Dice_i}]_{i=1}^{n}= 
\]
\[
 List_{Unc} = [Unc_{map_i}]_{i=1}^{n} 
\]

\textit{\textbf{Step 2: Generate ROI Uncertainty:}} 
\\
\hspace*{5mm} \textit{\textbf{Input:}} Test image $X_{test_i}$ and ground-truth mask $X_{GT_i}$. 
\newline
\hspace*{10mm} \textbullet \hspace*{1mm} $X_{test_i}$-lesion = $X_{GT_i}$, where `0' represents non-lesion pixels, `1' represents lesion pixels.
\newline
\hspace*{10mm} \textbullet \hspace*{1mm} $X_{test_i}$-non-lesion = $X_{test_i}$-lesion$^{C}$, complement of $X_{test_i}$-lesion, where `0' represents lesion pixels, `1' represents non-lesion pixels.
\newline
\hspace*{5mm} \textit{\textbf{Output:}} Uncertainty maps for specific regions computed as:
\newline
\hspace*{10mm} \textbullet \hspace*{1mm} $\bm{Unc_{i_L}}$ = $Unc_{test_i}$-lesion = ($X_{test_i}$-lesion $\odot$ $Unc_{map_i}$); $\bm{List_{T_{unc}}} = [mean(Unc_{i_L})]_{i=1}^{n}$.
\newline
\hspace*{10mm} \textbullet \hspace*{1mm} $\bm{Unc_{i_{NL}}}$ = $Unc_{test_i}$-non-lesion = ($X_{test_i}$-non-lesion $\odot$ $Unc_{map_i}$); $\bm{List_{NT_{unc}}} = [mean(Unc_{i_{NL}})]_{i=1}^{n}$. The symbol $\odot$  represents the Hadamard product.
\newline
\newline
\textit{\textbf{Step 3: Generate linear regression models for Dice Score Prediction:}} 
\\
\hspace*{5mm} \textit{\textbf{Input:}} Dependent variable $List_{Dice}$ = $Y_{Dice}$ (say) $= [y_{Dice_i}]_{i=1}^{n}$ for $n$ images, and independent variables:
\newline
\hspace*{10mm} \textbullet \hspace*{1mm} $List_{Unc}$ = $X_0$ (say) $= [x_{0i}]_{i=1}^{n}$ (overall uncertainty as shown in  Fig \ref{fig4} and \ref{fig3} column D).
\newline
\hspace*{10mm} \textbullet \hspace*{1mm} $List_{T_{unc}}$ = $X_1$ (say) $= [x_{1i}]_{i=1}^{n}$ (lesion tissue uncertainty as shown in Fig \ref{fig4} and \ref{fig3} column E)
\newline
\hspace*{10mm} \textbullet \hspace*{1mm} $List_{NT_{unc}}$ = $X_2$ (say) $= [x_{2i}]_{i=1}^{n}$ (non-lesion tissue uncertainty as shown in Fig \ref{fig4} and \ref{fig3} column F) -- that predict $Y_{Dice}$.
\newline
\hspace*{5mm} \textit{\textbf{Output:}} Model for $Y_{Dice}$ in terms of $X_1$ and $X_2$:
\begin{equation} \label{eq:1}
    Y_{Dice} = \alpha_0 + \alpha_1*X_1 + \alpha_2*X_2
\end{equation}
\vspace*{-2mm}
\hspace*{14mm}
\vspace*{-2mm}

\begin{equation} \label{eq:4}
\begin{aligned}
    \mbox{for individual ROIs of each image:} \\
    \begin{aligned}
    & (i) Y_{Dice} = \beta_0 + \beta_1*X_1
    & (ii) Y_{Dice} = \gamma_0 + \gamma_1*X_2
    \end{aligned}
\end{aligned}
\end{equation}

\hspace*{14mm} for the overall uncertainty, $X_0$:
\begin{equation} \label{eq:3}
    Y_{Dice} = \theta_a + \theta_b*X_0
\end{equation}

\hrule
\vspace{1mm}
\end{algorithm}

\section{Results} \label{sec:results_and_disc}

\subsection{Deep Learning Model Performance and Uncertainty Analysis} \label{sec:results_and_disc1}
This study focused on developing DLMs for skin lesion segmentation, leveraging uncertainty estimates to enhance trustworthiness and predict Dice coefficients for outputs using region based linear regression models. The utilization of MCD and BBP techniques for training two models, TFS and another based on TL, allowed for a comprehensive assessment of model uncertainty at the pixel level. Subsequently, the analysis was expanded to explore the combined effect of dropout and Bayes-by-backprop techniques within the TFS and TL frameworks. Five-fold cross-validation was implemented to segment skin lesions and cancer tissue to estimate segmentation uncertainties in the trained DLM models reported in this study. The confidence intervals (CI) were calculated using the empirical bootstrap method with n = 5,000 simulations.   Table \ref{tab1} shows that for the TFS model with dropout (tfs\_d), the reported Dice coefficient was 0.8809 (95\% CI: 0.8705 - 0.8802). For the TFS model with backpropagation (tfs\_bc), the reported Dice coefficient was 0.8162 (95\% CI: 0.8005 - 0.8250). When utilizing the TL model, the reported Dice coefficient for TL with dropout (tl\_d) were 0.8313 (95\% CI: 0.8205 - 0.8471), and  for TL with backpropagation (tl\_bc), the  Dice coefficient was 0.7702 (95\% CI: 0.7605 - 0.7824). In addition to the results from the entire test dataset, we also calculated deep learning model performance on three clinical classes individually. The models demonstrated distinct mean uncertainty values, for e.g. in melanoma tfs\_d had the highest mean uncertainty at 0.061448, while tl\_bc had the lowest at 0.03723. In the seborrhoeic class, tfs\_d also showed the highest mean uncertainty at 0.060842, with tl\_bc with the lowest at 0.037722. For nevus, the trend continued with tfs\_d resulting in the highest mean uncertainty at 0.052898 and tl\_bc the lowest at 0.026754.
The results indicated that both the TFS and TL models achieved high performance (Dice) in skin lesion and cancer segmentation using MCD and BBP (Table \ref{tab1}). The results also indicated that the tfs\_d model showed higher uncertainty across all metrics followed by the tl\_d model  (Table \ref{tab2}), but also achieved higher segmentation performance(Table \ref{tab1}). This heightened uncertainty can be attributed to the stochastic behavior of dropout, where units are randomly omitted during training introducing variability in predictions, resulting in higher uncertainty values. Training with backpropagation layers (tfs\_bc and tl\_bc) generally exhibited lower uncertainty compared to their dropout counterparts (Table \ref{tab2}), with corresponding lower segmentation performance (Table \ref{tab1}). We hypothesize that the dropout layers contribute to higher uncertainty metrics compared to backpropagation layers in both TL and TFS models, and therefore tfs\_d and tl\_d layers exhibit the higher uncertainty metrics  compared with tfs\_bc and tl\_bc models as shown in Table (\ref{tab2}. Therefore, the choice between dropout and backpropagation in  DLMs can be determined on the need for balanced performance and uncertainty management. The dropout method may be suitable for scenarios requiring both accuracy and awareness of uncertainty. Conversely, backpropagation provides stable predictions with lower uncertainty but may result in lower segmentation accuracy compared to dropout.

\begin{table}[htbp]
\centering
\caption{Median performance metrics for trained from scratch (TFS) and transfer learning (TL) deep learning models with Bayes backpropagation (bc) or Monte Carlo dropout (d) layers based on Dice score, AUROC, True Positive Rate (TPR), and False Positive Rate (FPR).}
\begin{tabular}{ p{1.3cm} p{1.3 cm} p{1.3cm} p{1.3cm} p{1.2cm} }
\hline
 & \multicolumn{2}{c}{\textbf{Performance}} \\
\hline
\textbf{Model} & \textbf{Dice  } & \textbf{AUROC} & \textbf{TPR} & \textbf{FPR} \\
\hline
tfs\_bc & 0.8162 & 0.8032 & 0.7829 & 0.0194 \\
tfs\_d & 0.8809 & 0.8662 & 0.7730 & 0.0008 \\
tl\_bc & 0.7702 & 0.7197 & 0.7068 & 0.0278 \\
tl\_d & 0.8313 & 0.8168 & 0.8353 & 0.0162 \\
\hline
\end{tabular}

\label{tab1}
\end{table}
\vspace{-0.3cm}

\begin{table}[htbp]
\centering
\caption{Uncertainty metrics for deep learning models trained from scratch (TFS) and transfer learning (TL) with Bayes backpropagation (BC) or Monte Carlo dropout (d).}
\begin{tabular}{ l c c c c }
\hline
 & \multicolumn{2}{c}{\textbf{Uncertainty}} \\
\hline
\textbf{Model} & \textbf{Maximum} & \textbf{Minimum} & \textbf{Mean} & \textbf{Std. dev.} \\
\hline
tfs\_bc  & 0.1094 & 0.0031 & 0.0128 & 0.0147 \\ 
tfs\_d & 0.1406 & 0.0063 & 0.0250 & 0.0420 \\
tl\_bc & 0.0976& 0.0029 & 0.0115 & 0.0145\\
tl\_d & 0.1305 & 0.0061 & 0.0159 & 0.0130 \\
\hline
\end{tabular}

\label{tab2}
\end{table}

\begin{figure}[h!]
\centering
\includegraphics[width=0.48\textwidth, height=0.21\textheight] {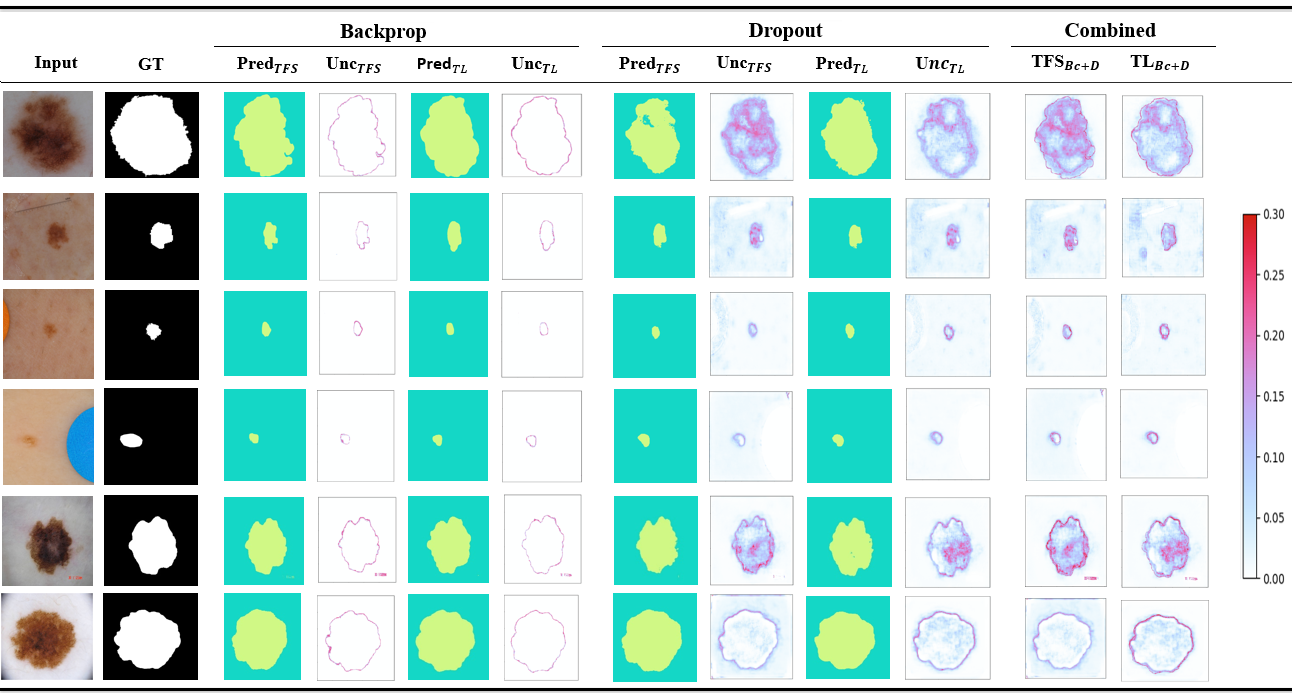}
\caption{Segmentation and uncertainty maps obtained from Bayes-by-backrop (Bc) and Monte Carlo dropout (MCD) augmented transfer learning (TL) and trained from scratch (TFS) deep learning models. GT - Ground Truth labels, Pred - Model prediction, Unc-Uncertainty.}
\label{fig2}
\end{figure}

Analyzing 2935 test images revealed that most of the uncertainty with the tl\_bc approach (n=2759 images) was primarily attributed to the skin lesion boundary, while 176 images exhibited uncertainty inside the lesion region (Fig \ref{fig2}). Similarly, using the tfs\_bc approach on the same set of test images, only 253 images demonstrated uncertainty within the lesion region, with the majority (2682 images) showing uncertainty localized on the boundaries (Fig \ref{fig2}). These findings indicate that the MCD Bc (tl\_bc, tfs\_bc) used in this study prioritizes the detection of uncertainty in boundary segmentation. Conversely, with the tl\_d approach, 2924 images exhibited segmentation uncertainty inside the lesion region, and only 11 images had boundary pixels with uncertainty. Similarly, with the tfs\_d approach, uncertainty was centered in the interior of the lesion regions in 2927 images, with only eight images emphasizing the boundary. Thus, the MCD approach with both TFS and TL in DLMs shows the highest uncertainty inside skin lesions and lower uncertainty in other regions of the RGB images.

This also suggests that the MCD approach exhibits lower performance when predicting within lesion tissue regions but performs well in other regions.

\subsubsection{Performance On Prediction Of Nevus Skin Tissue} \label{sec:results_and_disc3a} 
 From  lesion containing images $\alpha_0$ = 0.8449, $\alpha_1$ = -1.1932 and $\alpha_2$ = -1.5108 underscore the predictive behavior with low RMSE values ($\leq 0.16$) indicating accurate predictions across all four linear regression models. The mean Dice value of 0.837642 further indicated the DLM and linear regression model efficacy in delineating nevus skin tissue pixels. After evaluating the coefficients and ROI-based uncertainties in Table \ref{combined_results_horizontal} column 3, the regression model exhibited varying levels of certainty in different regions of interest. These insights can assist with the interpretation of predictive performance and the reliability of DLMs in segmenting nevus skin lesions.

\subsubsection{Performance On Prediction Of Melanoma Skin Lesions} \label{sec:results_and_disc3b}
 Table \ref{combined_results_horizontal} column 4 shows results for melanoma skin lesions detailing the coefficients and values associated with ROI-based uncertainty. From these images $\alpha_0$ = 0.9644, $\alpha_1$ = -8.6780 and $\alpha_2$ = -3.0881 were calculated. The RMSE values for all four linear regression models were $\leq 0.16$ and the mean Dice value were 0.822958.

\subsubsection{Performance On Prediction Of Seborrhoeic Keratosis Skin Tissue} \label{sec:results_and_disc3c}
 Table \ref{combined_results_horizontal} column 5 shows a detailed breakdown of the coefficients and associated values that explain the uncertainty within specific ROIs for seborrhoeic keratosis skin tissue pixels. From these lesion containing images, $\alpha_0$ = 0.8835, $\alpha_1$ = -8.4215 and $\alpha_2$ = -4.7513 were calculated. The low RMSE values ( $\leq 0.19$) obtained across all four linear models indicated that the models reported high segmentation performance in predicting the lesion presence in these images. Additionally, the mean Dice coefficient of 0.845437 reflects the efficacy of the models in delineating seborrhoeic keratosis skin tissue pixels.
 \vspace{-0.3cm}
\begin{table}[htbp]
  \caption{Analysis of coefficients and correlation in skin tissue classes. ROI: clinical regions of interest in images. The values of $\beta_0$, $\beta_1$, $\gamma_0$, $\gamma_1$, $\theta_a$, and $\theta_b$ are linear regression coefficients. $\rho$ represents the Spearman's correlation coefficient.}                                   \resizebox{\columnwidth}{!}{%
    \begin{tabular}{l l c c c}
    \hline
    \textbf{ROI-based Uncertainty} & \textbf{Coefficients} & \textbf{Nevus} & \textbf{Melanoma} & \textbf{Seborrhoeic} \\
    \hline
    \multirow{3}{*}{Lesion uncertainty} 
        & $\beta_0$  & 0.8350 & 0.9135 & 0.8038 \\
        & $\beta_1$  & -1.5644 & -7.7102 & -6.3832 \\
        & $\rho$  & -0.0243 & -0.486 & -0.3042 \\
    \hline
    \multirow{3}{*}{Non-lesion uncertainty} 
        & $\gamma_0$  & 0.8433 & 0.8311 & 0.8049 \\
        & $\gamma_1$  & -1.5579 & -0.5570 & -3.1506 \\
        & $\rho$  & 0.1275 & 0.0008 & -0.1538 \\
    \hline
    \multirow{3}{*}{Overall uncertainty} 
        & $\theta_a$  & 0.8449 & 0.9643 & 0.8769 \\
        & $\theta_b$  & -1.4596 & 5.7393 & -5.6249 \\
        & $\rho$  & -0.8496 & -0.5556 & -0.4164 \\
    \hline
    \end{tabular}%
    }
  
    \label{combined_results_horizontal}
\end{table}

\vspace{-0.2cm}

\subsubsection{Predictive capabilities of the proposed linear models} \label{subsection:predictive performance}
Eq.~\ref{eq:1} was further expanded with the class information for an input image as categorical variable. Thus, Eq.~\ref{eq:1} can be re-written for this case as $Y_{Dice} = (\alpha_0 + \alpha_1*X_1 + \alpha_2*X_2 + \phi_1*C_1 + \phi_2*C_2 + \phi_3*C_3)$, where $C_1$, $C_2$ and $C_3$ (that can take either a value of 0 or 1 based on presence or absence of the class information) correspond to the melanoma, nevus and seborrhoeic keratosis diagnosis classes. Solving for the values of these coefficients based on the entire test set of 2,935 images, $\alpha_0$ = 0.8852, $\alpha_1$ = -7.5594, $\alpha_2$ = -2.6110, $\phi_1$ = 0.06, $\phi_2$ = -0.0229 and $\phi_3$ = -0.0372 were calculated. A low RMSE of 0.1543 was calculated for this linear regression model. A positive coefficient for all classes indicated high segmentation performance. For all derived coefficients in each of the three clinical classes, negative coefficient ($\alpha_1$, $\alpha_2$, $\beta_1$, $\gamma_1$ and $\theta_b$) values indicated that uncertainty in linear regression models predictions decreased its performance. From the coefficients derived for Eq.~\ref{eq:1}, we further note the importance of each clinical region as well as the degree to which a clinical diagnosis class affects the trained model's overall performance -- a lesser (more negative) value of the coefficient implying a greater contribution in reducing Dice. For the images belonging to the nevus class, the non-lesion region uncertainties had the highest coefficient value (weight) relative to the other region. These results also indicated that linear regression models performance was most affected (reduced) by the uncertainty of predictions of non-lesion region, followed by the lesion-tissue regions. For the melanoma and seborrhoeic keratosis images, lesion tissue region uncertainties contributed more substantially than pixels without any lesion tissue.

\subsection{Comparative Qualitative Evaluation of Deep Learning Models} \label{sec:results_and_disc2}

Monte Carlo dropout consistently outperformed backpropagation DLMs in skin lesion and cancer segmentation (Table \ref{tab1}). Consequently, the qualitative evaluation focused on MCD in TFS and TL models. Visualizations of segmentation masks and uncertainty maps (Figs. \ref{fig4} and \ref{fig3}) revealed MCD model performance across clinical scenarios. The study also examined the impact of training approach (TFS vs. TL) and MCD utilization on model performance. Figures \ref{fig4} and \ref{fig3} display RGB skin dermoscopy images with model output segmentation masks and uncertainty heatmaps. Model predictions and overall uncertainty maps are shown in columns C and D, while region-based uncertainty maps, $\bm{Unc_{i_L}}$ and $\bm{Unc_{i_{NL}}}$, are in columns E and F (Figs. \ref{fig4} and \ref{fig3}).

\subsubsection{Qualitative Evaluation of the Monte Carlo Dropout Trained From Scratch Deep Learning Model}

\begin{figure}[h!]
\centering  
\includegraphics[width=0.48\textwidth, height=0.23\textheight]{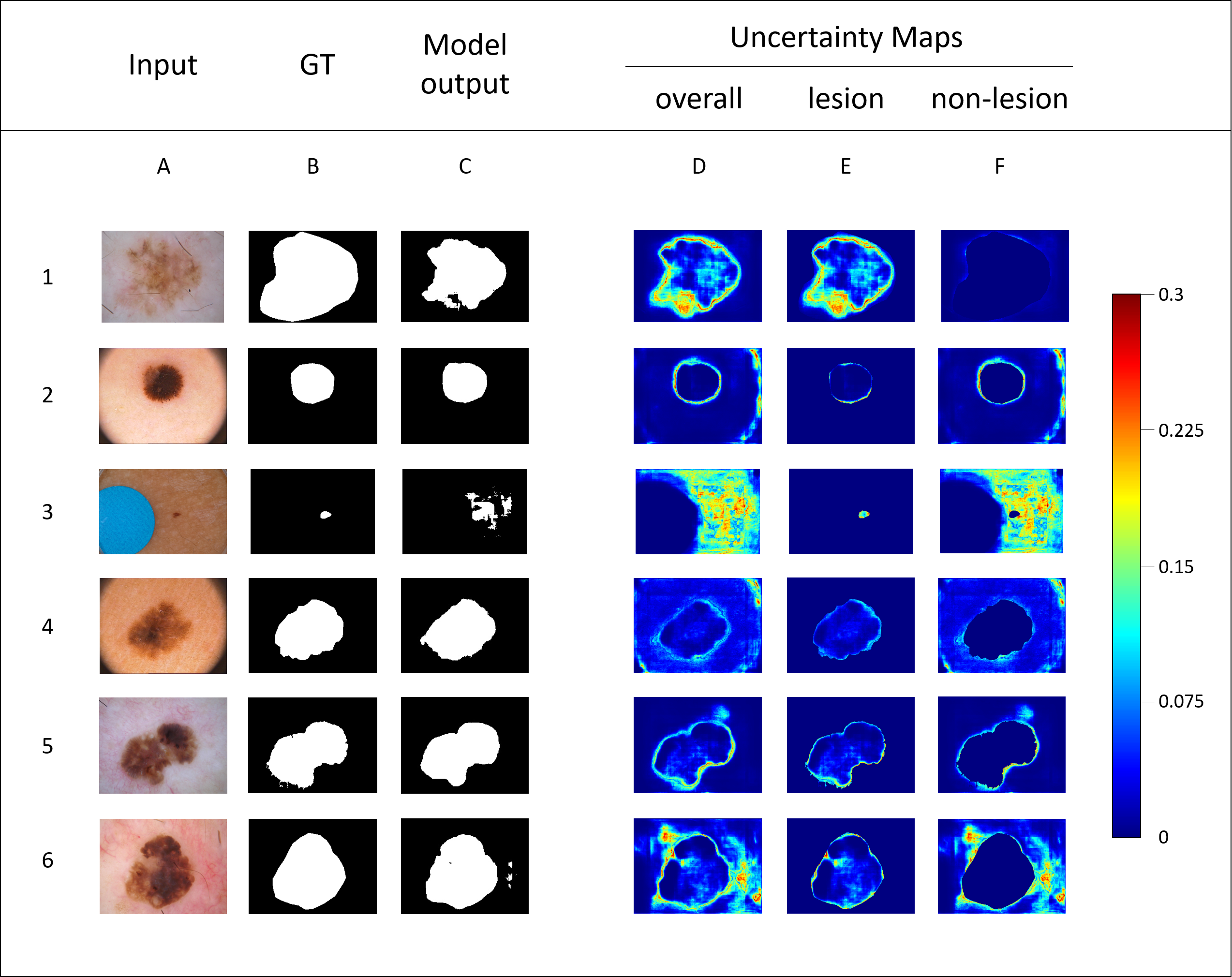}

\caption{\footnotesize{Visualization of skin cancer segmentation and uncertainty by a Monte Carlo dropout (MCD) model trained from scratch on dermoscopy images. \textbf{Columns left to right:} \textbf{A} Input RGB image; \textbf{B} Clinical ground-truth binary mask; \textbf{C} MCD segmentation binary mask; \textbf{D} MCD uncertainty map for segmentation; \textbf{E} MCD uncertainty map for lesion regions; \textbf{F} MCD uncertainty map for non-lesion regions. The color bar shows model uncertainty, with red indicating high uncertainty and blue indicating low uncertainty (0). 'GT' stands for clinical ground-truth.}}
\label{fig4}
\end{figure}

\textbf{Figure~\ref{fig4}} shows six (two from each clinical class) representative skin dermoscopy images from the test data and MCD model output segmentation masks and corresponding uncertainty heatmaps from  trained TFS model (tfs\_d). Images 1 and 2 in Fig.~\ref{fig4} belong to the melanoma class, images 3 and 4 to the nevus class and images 5 and 6 to seborrhoeic keratosis. Predictions for images 2, 4, 5 and 6 show almost no false negative and no or very low false positive regions indicating good segmentation accuracy. Image 1 exhibited the highest false negative while image 3 showed the highest false positive regions. For image 1, the false negative regions were prominent where the tfs\_d model showed the highest uncertainties -- this could have been a result of the model being unable to differentiate between lighter colored pixels tissues which are part of the lesion ground-truth and the background that contains non-lesion skin tissue which are of similar coloration. Such lighter pixels color in  skin lesion and tumor signatures were not effectively captured by the tfs\_d possibly due to the sparsity of such examples in the data set and is indicated by the high uncertainty values for the region (see Fig.~\ref{fig4} image 1D and 1E). 

Figure ~\ref{fig4} image 1F shows low uncertainty for pixels in non-lesion regions, indicating superior performance. The uncertainty maps (column D) and lesion uncertainty maps (column E) demonstrate that the tfs\_d model effectively learned the lesion regions and exhibited less uncertainty in non-lesion regions (column F). Most detected uncertainties were within the non-lesion region, at the lesion and non-lesion boundary, or at skin tissue and non-tissue boundaries (images 2 and 4). This indicates that the tfs\_d model accurately learned lesion regions but showed uncertainty at non-lesion skin tissue, non-tissue junctions, and tissue intersections.

In images 3 and 6, uncertainties in non-lesion regions suggest that the tfs\_d model may inaccurately segment darker non-lesion skin tones with lesions, especially in image 3 where the lesion color closely matches the non-lesion color. Additionally, in image 3, the model effectively ignored a blue artifact, as indicated by no uncertainty in that region. Visualizing these ROI-based uncertainty maps (Fig. ~\ref{fig4}) helps identify high-uncertainty regions and make informed decisions on model output. High uncertainties in the lesion region of image 1 suggest potential lower segmentation performance within lesion regions during inference, while similar uncertainties in image 6 for both lesion and non-lesion regions indicate acceptable predictions when differential diagnosis depends on disease pixels.

\subsubsection{Qualitative evaluation of the  Monte Carlo dropout transfer learning deep learning model}

\begin{figure}[h!]
\centering
\includegraphics[width=0.48\textwidth, height=0.20\textheight]{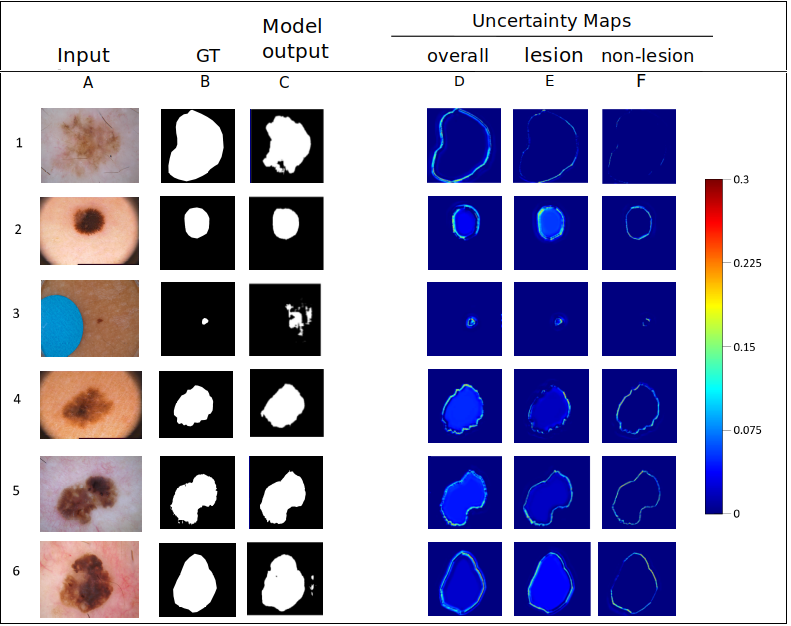}
\caption{\footnotesize{Visualization of skin cancer segmentation and uncertainty by a Monte Carlo dropout (MCD) model trained with transfer learning on dermoscopy images. \textbf{Left to right columns:} \textbf{A} Input RGB image; \textbf{B} Clinical ground-truth binary mask; \textbf{C} MCD segmentation binary mask; \textbf{D} MCD uncertainty map for segmentation; \textbf{E} MCD uncertainty map for lesion regions; \textbf{F} MCD uncertainty map for non-lesion regions. The color bar shows model uncertainty, with red indicating high uncertainty and blue indicating low uncertainty (0). 'GT' stands for clinical ground-truth.}}
\label{fig3}
\end{figure}

\textbf{Figure.~\ref{fig3}} shows six representative skin dermoscopy images from the test data along with the output segmentation masks and corresponding uncertainty heatmaps from the trained Monte Carlo dropout transfer learning deep learning (tl\_d) model. Images 1 and 2 in Fig.~\ref{fig3} belong to the melanoma class, images 3 and 4 to the nevus class, and images 5 and 6 to the seborrheic keratosis. Similar to the tfs\_d  evaluation, we observed that predictions for images 1, 2, 4, 5, and 6 exhibited almost no false negatives and no or very low false positive regions, indicating good segmentation accuracy. However, there were differences in the uncertainty patterns and performance analysis compared to the tfs\_d. For instance, in image 1 (Fig. ~\ref{fig4} and ~\ref{fig3}) the tfs\_d model exhibited higher uncertainties in the lesion region compared to the tl\_d model, indicating potential challenges in accurately segmenting melanoma lesions during inference. Furthermore images 3 and 6 (Fig. ~\ref{fig4} and ~\ref{fig3}) showed lower uncertainties in the non-lesion regions in tl\_d model compared to the tfs\_d, suggesting improved performance of the tl\_d model in distinguishing between lesion and non-lesion skin tissue in these cases.

Visualizing the ROI-based uncertainty maps (Fig.~\ref{fig3}) helped identify regions of high uncertainties, informing decisions about the model’s output. In image 1, high uncertainties were observed at the boundaries, indicating potential improvement areas for the tl\_d model. Unlike the tfs\_d model, the tl\_d model showed minimal uncertainties in both lesion and non-lesion regions, suggesting more reliable predictions for differential diagnosis reliant on disease pixels. Overall uncertainty maps (column D) and lesion uncertainty maps (column E) indicated that the tl\_d model learned lesion regions well, exhibiting less uncertainty compared to tfs\_d (column F) (Fig. ~\ref{fig4} and ~\ref{fig3}). Most detected uncertainties were within the lesion and non-lesion boundary regions or at the junctions between skin tissue and non-tissue. This demonstrated that the tl\_d model had effectively learned lesion regions and showed less uncertainty at non-lesion skin tissue and tissue intersections. Both models showed strengths in segmentation accuracy and learning lesion regions but faced challenges with specific lesion types. The tl\_d model was better in distinguishing between lesion and non-lesion tissues, providing more reliable predictions for disease pixels compared to tfs\_d (Fig. ~\ref{fig4} and ~\ref{fig3}). For the tl\_d model, uncertainties were prominent around the edges where lesion tissue pixels meet non-lesion or background tissue pixels, while for the tfs\_d model, uncertainties were prominent within lesion tissues. The tfs\_d model was less optimal in predicting non-lesion tissue regions, possibly due to insufficient examples with clear boundary demarcations and differential features in the dataset.

\vspace{-0.1cm}

\section{Discussion}
In this study, the performance of deep learning models (DLMs) using TFS and TL with Bayes-by-Backprop (Bc) and Monte Carlo Dropout (MCD) techniques for skin cancer segmentation was examined (Fig. \ref{fig2}). The models employing TFS or TL with Bayes-by-Backprop (tfs\_bc, tl\_bc) showed higher segmentation uncertainty concentrated on the boundaries of the skin lesion, indicating lower uncertainty within the lesion region. These boundaries typically represent areas where skin tumor tissue pixels intersect with non-lesion or background tissue pixels. In contrast, TFS or TL with Monte Carlo Dropout (tfs\_d, tl\_d)showed more prevalent uncertainties within the tumor tissues compared to the DLM Bc model (see Fig. \ref{fig2}).

The combination of DLMs Bc (tfs\_bc, tl\_bc) and DLMs MCD (tfs\_d, tl\_d) techniques involved integrating predictions from each method. This approach leveraged the strengths of both techniques to improve overall performance and generalization. These results show that combining DLM Bc and DLM MCD predictions within the TFS and TL frameworks reduced uncertainty and provided more robust predictions, improving segmentation accuracy with low uncertainties (Table \ref{tab1} and \ref{tab2}). 

This study also developed a novel method to estimate uncertainty in DLM predictions for skin lesion segmentation. By correlating uncertainty with Dice scores using linear regression, the proposed algorithm enhances performance estimation and interpretability through the generation of uncertainty maps.

Quantitative analysis through five-fold cross-validation revealed significant correlations between predictive performance, uncertainty measures, and ground-truth segmentation masks. Qualitative evaluation validated the effectiveness of the trained models, reporting accurate segmentation with minimal false positive and false negative pixels. The study also offered specific guidelines for choosing between TFS and TL approaches in skin lesion and cancer image segmentation tasks by DLMs. High performance metrics in AUROC, Dice coefficient, and TPR were achieved by employing TFS with dropout layers, delivering superior performance and robust uncertainty estimates (Table \ref{tab1} and \ref{tab2}). Using TL DLM with dropout layers balanced performance and uncertainty, making it suitable for scenarios requiring both accuracy and uncertainty awareness. While TFS with backpropagation layers provided stable predictions with lower uncertainty estimations, albeit with potential trade-offs in lower segmentation performance(Table \ref{tab1} and \ref{tab2}). Thus, integrating both dropout and backpropagation techniques within TFS and TL frameworks results in a comprehensive uncertainty reduction strategy for enhancing model robustness and precision for skin lesion segmentation (Fig. \ref{fig2}).

The trained deep learning models from this study accurately align with ground-truth data and can be adapted and fine-tuned using Algorithm 1 for small amounts of unlabelled, out-of-distribution (OoD) data to minimize false positives and maintain high performance levels. When tested with OoD data lacking true ground-truth, the study employs historical performance metrics such as Dice scores and AUROC (Table \ref{tab1} and \ref{tab2}) to evaluate model effectiveness, utilizing Equation ~\ref{eq:3} where appropriate. Thus, Algorithm 1 is versatile and promising for working with other skin imaging datasets for improving segmentation tasks.

\paragraph{Limitations}
While the performance metrics of the framework, approach, and models in this study are promising, it is important to acknowledge that the calculated performance could vary for other practitioners. This variability can be attributed to several factors unique to their study, including the complexity of the model architecture, data preprocessing techniques, refinement of feature selection processes, and the choice of pipeline.

\section{Conclusion}

This study analyzes the relationship between prediction uncertainty and the Dice score of whole slide image segmentation. It demonstrates that clinical region-based linear models can predict DLM performance (Dice) using individual or combined uncertainty estimations (Equations: ~\ref{eq:1}-~\ref{eq:3}). Unlike other studies that focus only on overall uncertainty measures, this study decouples clinical and region-based uncertainty estimates for skin lesions and cancer signatures. Significant correlations (p $<$ 0.05) between overall and ROI-based uncertainties with Dice scores were established using Spearman’s rank correlation. High negative values indicated a strong negative correlation between uncertainty and model performance, supported by linear regression coefficients, most of which were negative. Low RMSE indicated a good fit for these results. Equation ~\ref{eq:1} predicts performance based on all regions simultaneously, identifying regions contributing to greater uncertainty and poorer performance. Algorithm ~\ref{seg_algo1} estimates model performance unsupervised, generating uncertainty maps and mean uncertainty values. Clinical ROI-based uncertainty estimates and predicted Dice scores can inform real-world clinical decisions. Future research will evaluate models across diverse skin and other organs for multi-class segmentation, integrate domain knowledge, and simulate aleatoric and epistemic noise during training for further refinement in clinical and machine learning outcomes.

\vspace{-1mm}
\section{Data, Code and Model Availability}

All  resources, results, data, code, and models, are deposited in a GitHub repository \url{https://github.com/Prof-Pratik-Shah-Lab-UCI/skin_project_2024}, and will be made available upon request to the corresponding author.

\bibliographystyle{IEEEtran}

\bibliography{sample}

\end{document}